\documentclass[aps,amsmath,twocolumn,amssymb,floatfix,showpacs,superscriptaddress,nofootinbib,longbibliography]{revtex4-1}
\usepackage{MnSymbol}
\usepackage{braket}
\usepackage[dvipsnames]{xcolor}
\usepackage{float}
\usepackage{subfigure}
\usepackage{tikz}
\usepackage[colorlinks=true,linktoc=page,linkcolor=red,citecolor=blue,urlcolor=purple]{hyperref}

\mathchardef\mhyphen="2D 

\newcommand{\ie}{{ i.e.,\,\,}}
\newcommand{\eg}{{e.g.,~}}

\newcommand\bea{\begin{eqnarray}}
\newcommand\eea{\end{eqnarray}}
\newcommand\beq{\begin{equation}}  
\newcommand\eeq{\end{equation}}

\newcommand{\non}{\nonumber}  
\usepackage[normalem]{ulem}
\definecolor{lime}{HTML}{A6CE39}
\usepackage{sidecap,tikz}
\DeclareRobustCommand{\orcidicon}{\hspace{-1.0mm}
	\begin{tikzpicture}
		\draw[lime, fill=lime] (0.0,0.0) 
		circle [radius=0.15] 
		node[white] {{\fontfamily{qag}\selectfont \tiny \,ID}};
		\draw[white, fill=white] (-0.0525,0.095) 
		circle [radius=0.007];
	\end{tikzpicture}
	\hspace{-3.0mm}
}
\foreach \x in {A, ..., Z}{\expandafter\xdef\csname orcid\x\endcsname{\noexpand\href{https://orcid.org/\csname orcidauthor\x\endcsname}{\noexpand\orcidicon}}
}

\AtBeginDocument{%
	\newwrite\bibnotes
	\def\bibnotesext{Notes.bib}
	\immediate\openout\bibnotes=\jobname\bibnotesext
	\immediate\write\bibnotes{@CONTROL{REVTEX41Control}}
	\immediate\write\bibnotes{@CONTROL{%
			apsrev41Control,author="08",editor="1",pages="1",title="1",year="1"}}
	\if@filesw
	\immediate\write\@auxout{\string\citation{apsrev41Control}}%
	\fi
}%

\begin{document}

\title{Generation and time evolution of anomalous Floquet Majorana flat edge modes in two-dimensional noncolinear magnet-superconductor heterostructures}  

\author{Kamalesh Bera\orcidA{}}
\email{kamalesh.bera@iopb.res.in}
\affiliation{Institute of Physics, Sachivalaya Marg, Bhubaneswar-751005, India}
\affiliation{Homi Bhabha National Institute, Training School Complex, Anushakti Nagar, Mumbai 400094, India}

\author{Priyanka Mohan\orcidB{}}
\email{priya11198@gmail.com}
\affiliation{Dolat Capital Market Pvt Ltd, India}

\author{Arijit Saha\orcidC{}}
\email{arijit@iopb.res.in}
\affiliation{Institute of Physics, Sachivalaya Marg, Bhubaneswar-751005, India}
\affiliation{Homi Bhabha National Institute, Training School Complex, Anushakti Nagar, Mumbai 400094, India}

\begin{abstract}
We theoretically investigate the realization of gapless Floquet topological superconducting phases in a two-dimensional magnet-superconductor heterostructure 
(2D Shiba lattice) in the presence of a harmonic drive implemented in the chemical potential. Employing a real-space tight-binding model, we obtain both the regular $0$- and anomalous $\pi$-Floquet Majorana flat edge modes (FMFEMs) in the quasi-energy spectrum. We also study the real-time evolution of the FMFEMs 
and analyze their local density of states in the presence of such a periodic drive. The topological characterization is performed using the winding number, exploiting the chiral symmetry of the equivalent bulk effective momentum-space Hamiltonian. This is also supported by the corresponding edge state spectra. Furthermore, we employ the Brillouin-Wigner (BW) and Floquet perturbation theory (FPT) to gain analytical insight into the problem. We compare our exact (numerical), BW, and FPT results in terms of the quasi-energy spectra obtained across different frequency regimes. We find good agreement between the exact numerical, BW, and FPT results in the higher-frequency and high-amplitude domain, particularly close to the $0$-quasi-energy modes. 
\end{abstract}

\maketitle

\section{Introduction}

In recent years, the investigation of topological superconductivity (TSC) in different heterostructures has become intense research interest, as it provides a platform to realize Majorana bound states (MBSs), which can be useful for building fault-tolerant quantum computers~\cite{KITAEV20032, RMP_Nayak}. Although the idea of MBSs was first introduced by Kitaev through a one-dimensional (1D) spinless $p$-wave superconducting chain~\cite{Kitaev2001}, the rarity of $p$-wave superconductivity in nature makes its direct practical realization challenging. To overcome this, various theoretical proposals have been put forward for an indirect realization of MBSs. One of the most promising platform among them is the consideration of a semiconducting nanowire with strong spin-orbit coupling and a Zeeman field placed in close proximity to a conventional $s$-wave superconductor~\cite{RNW_thr1, RNW_thr2, RNW_thr3, RNW_thr4}. Despite intense experimental efforts, there have been reports claiming the indirect observation of MBSs~\cite{RNW_expt1, RNW_expt2}, but none have established it's smoking gun signals till date~\cite{RNW_expt3}. In recent times, the experimental observation~\cite{YSR_expt} of Yu-Shiba-Rusinov (YSR) states inside the bulk superconducting gap, due to the presence of magnetic impurities in them~\cite{YSR_theory}, has opened another avenue for engineering the 1D $p$-wave Kitaev chain physics. Proposals based on 1D helical spin chains~\cite{HSC_thr1,HSC_thr2,HSC_thr3,Imp_chn1,Imp_chn2,Imp_chn3} have been put forward, and later experimental signatures signaling MBSs were probed in such systems through scanning tunneling microscopy (STM) measurements~\cite{Imp_chn_expt1,Imp_chn_expt2}.

Apart from the 1D models, various other setups have also been proposed in higher dimension to realize MBSs. For \eg the surface states of three-dimensional (3D) topological insulator (TI) in close proximity to an $s$-wave superconductor and ferromagnetic insulator~\cite{FuKane_2008}, quantum Hall and anomalous Hall systems~\cite{TSC_QH_Zhang, TSC_graphene1, TSC_graphene2} in proximity to an $s$-wave superconductor, along with the two-dimensional (2D) versions of the previously discussed 1D models are worth mentioning ~\cite{2D_SOC_Bz1,2D_SOC_Bz2,2D_SOC_Bz3,2D_spin_tex1,2D_spin_tex2,2D_spin_tex3,2D_spin_tex4}. Interestingly, in 2D, unlike in 1D, TSC in Kitaev Model can appear with different types of $p$-wave pairings: gapped ``$p_{x} + ip_{y}$''-type pairing hosting dispersive Majorana edge modes~\cite{2D_SOC_Bz1, 2D_SOC_Bz2,2D_spin_tex2} and gapless ``$p_{x} + p_{y}$''-type pairing anchoring flat Majorana edge modes~\cite{2D_gapless_KM1,2D_gapless_KM2,2D_SOC_Bz3}. Very recently, the appearance of such Majorana flat edge modes has been established in 2D non-collinear magnetic texture platform ~\cite{2D_spin_tex1,2D_spin_tex4,2D_spin_tex5,2D_spin_tex6}.

On the other hand, \emph{Floquet engineering}~\cite{Floq_topo_rev1,Floq_topo_rev2} has emerged as a powerful tool for tailoring quantum materials and realizing novel phases of matter those are absent in the static systems. By applying a time-periodic drive, it is possible to induce topologically non-trivial phases starting from an otherwise trivial system~\cite{Floq_topo_thry1,Floq_topo_thry2,Floq_topo_thry3,Floq_topo_thry4,Floq_topo_thry5,Floq_topo_thry6,Floq_topo_thry7,expt_FLQ1,expt_FLQ2,expt_FLQ4,expt_FLQ5,expt_FLQ6}. Remarkably, periodically driven systems can host not only the conventional zero-energy modes but also anomalous $\pi$-modes, which have no static counterparts~\cite{Floq_pi1,Floq_pi2,expt_FLQ3,Floq_pi3, Floq_pi4}. On the other hand, Floquet-induced TSC remains a comparatively less explored area in higher dimensions. Although a variety of 1D platforms, ranging from cold atomic systems, Rashba nanowires to helical Shiba chain, have been theoretically investigated in this context~\cite{1D_Floq_TSC1,1D_Floq_TSC2,1D_Floq_TSC3,1D_Floq_TSC4,1D_Floq_TSC5,1D_Floq_TSC6} and emergence of Floquet Majorana modes ($0$,$\pi$) have been proposed. In recent times, some of these studies have been extended to two-dimensional (2D) systems, where Floquet topological superconducting (FTSC) phases (both the first and higher order topology) have also been explored~\cite{2D_Floq_TSC1,2D_Floq_TSC2,2D_Floq_TSC3,2D_Floq_TSC4,2D_Floq_TSC5,2D_Floq_TSC6,2D_Floq_TSC7}.

In this article, we address the following intriguing questions: (a) Similar to the static system, which hosts Majorana flat edge modes (MFEMs), can we engineer their anomalous counterparts through Floquet driving 
in non-colinear magnet-superconductor (SC) heterostructure? (b) If so, then how do these modes evolve in time and how can they be topologically characterized? (c) Can perturbative approaches be employed to analytically understand these emergent driven phases?

To answer these questions, in this work, we consider a 2D non-collinear spin texture deposited on the surface of a conventional $s$-wave superconductor. The chemical potential is considered to be driven by a time-periodic modulation. The corresponding static model hosts $0$-MFEMs for suitable parameter 
regimes~\cite{2D_spin_tex4}. In the first part of our study, using a real-space tight-binding model, we analyze the quasi-energy spectrum obtained from the periodized evolution operator. Starting from the topological phase of the static system, we subject that to periodic driving at two distinct frequencies. In the low-frequency regime, we observe the emergence of anomalous $\pi$- Floquet Majorana flat edge modes (FMFEMs), which have no static counterparts. In contrast, at intermediate driving frequency, the system simultaneously hosts both $0$- and $\pi$-FMFEMs. To further analyse these modes, we compute the local density of states (LDOS). In the low-frequency regime, the LDOS corresponding to the $\pi$-modes reveals clear signatures of edge localization. Similarly, at intermediate frequency, the LDOS associated with the $0$-modes also exhibits 1D flat edge states. We then investigate the time evolution of the quasi-energy spectrum within 
a single driving period and 
present the LDOS at three different time instants, demonstrating the time evolution of both $0$- and $\pi$-modes. To evaluate the topological invariant, we construct an  effective momentum-space Hamiltonian starting from the real-space Hamiltonian. Further, we construct an edge Hamiltonian and calculate the quasi-energy edge spectra, which exhibit $0$- and $\pi$-FMFEMs, consistent with our real-space tight-binding results. Utilizing the chiral symmetry of the bulk Hamiltonian and extending it to the periodically driven case, we compute the dynamical winding number to characterise these Floquet topological phases. We find that the winding number assumes finite integer values in the parameter regions where $0$- and $\pi$-modes appear in the quasi-energy edge spectra. 
Finally, we perform a perturbative analysis to gain analytical insight into our numerical results. In particular, we employ both the Brillouin-Wigner (BW) and Floquet perturbation theory (FPT) to compute the quasi-energy spectra and compare those results with the exact numerical results obtained from the real-space tight-binding model across the three driving frequency regimes. We find that while these perturbative approaches fail in the low-frequency regime, they exhibit good agreement with the numerical results in both the intermediate- and high-frequency regimes for the $0$-FMFEMs.

The remainder of the article is structured as follows. In Sec.~\ref{Sec:II}, we present our model Hamiltonian, the driving protocol, and the underlying formalism. Sec.~\ref{Sec:III} contains the numerical results obtained using the Floquet operator based on real-space tight-binding model Hamiltonian. 
In Sec.~\ref{Sec:IV}, we analyze the topological properties of the dynamical $0$- and $\pi$-FMFEMs by computing the relevant topological invariant and examining their corresponding boundary edge spectra. The analytical results based on BW and FPT, together with their comparison to the exact numerical results, are discussed in Sec.~\ref{Sec:V}. Finally, we summarize and conclude our findings in Sec.~\ref{Sec:VI}. 

\section{Model and Method}\label{Sec:II}
In this section, we introduce our model Hamiltonian, driving protocol, and the formalism used to deal with the driven 2D magnet-SC heterostructure.

\begin{figure}[h]
	\centering
	\subfigure{\includegraphics[width=0.48\textwidth]{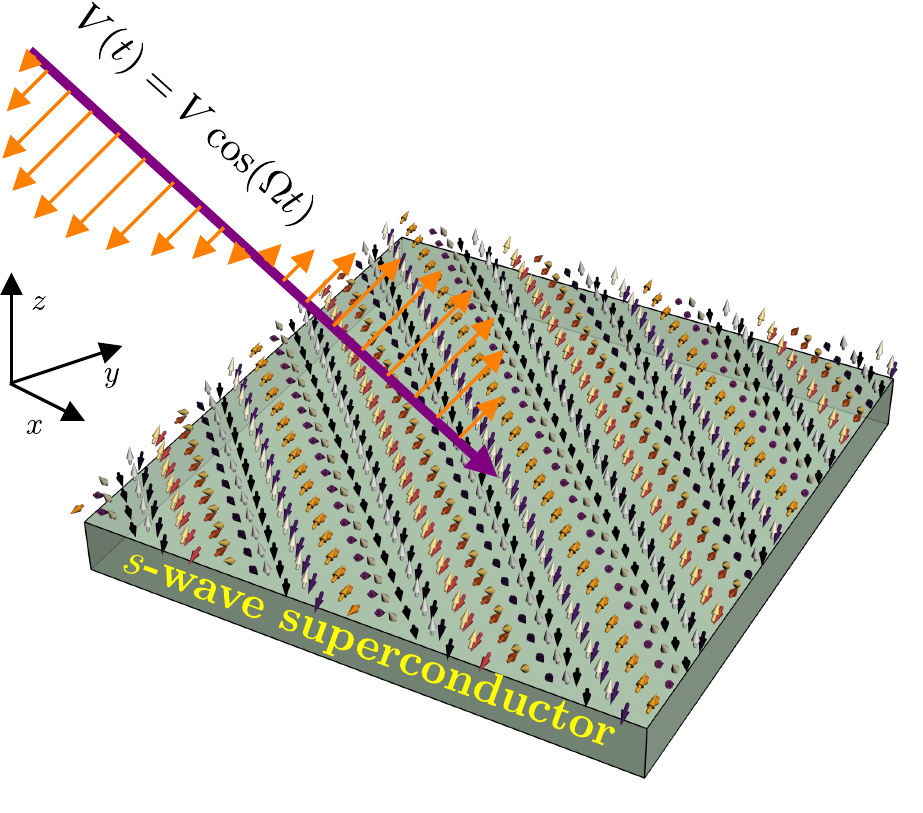}}
	\caption{Schematic representation of a 2D non-collinear spin texture decorated on a square lattice and deposited on the surface of a common $s$-wave SC. 
		A time-periodic harmonic drive is applied to the onsite chemical potential to realize $0$- and anomalous $\pi$-FMFEMs.
	}
	\label{schematic}
\end{figure}
\subsection{Model Hamiltonian}

We consider a hybrid system consisting of a 2D spin texture placed on the surface of a regular $s$-wave SC~(see Fig.~\ref{schematic} for a schematic representation). The Bogoliubov-de-Gennes (BdG) Hamiltonian for such a system in real-space continuum is given by $\mathcal{H} = \int d\mathbf{r} \, \Psi^{\dagger}(\mathbf{r}) H_{\text{BdG}} \Psi(\mathbf{r})$, where $\Psi(\mathbf{r}) = (c_{\mathbf{r} \uparrow}, c_{\mathbf{r} \downarrow}, c_{\mathbf{r} \downarrow}^{\dagger}, -c_{\mathbf{r} \uparrow}^{\dagger})^{\text{T}}$ is the Nambu spinor. Here, $c_{\mathbf{r} \uparrow}$ represents the annihilation operator of electron at position $\mathbf{r}$ with spin-up. The Hamiltonian in the first-quantized form can be written as,
\begin{eqnarray}\label{Conti_real_static_Ham}
	H_{0} &=& - \frac{\hbar^{2}}{2m} \boldsymbol{\nabla}^{2} \, \tau_{z} - \mu \, \tau_{z}
	- J \, \mathbf{S}(\mathbf{r}) \cdot \boldsymbol{\sigma}
	+ \Delta_{0} \, \tau_{x}\ ,
\end{eqnarray} 

Here, $\mu$, $J$, and $\Delta_{0}$ denote the chemical potential, the exchange coupling between the magnetic impurities and the superconducting electrons (from locally broken Cooper pairs), and the superconducting pairing strength, respectively. The matrices $\sigma$ and $\tau$ represent the Pauli matrices acting on the spin and particle-hole subspaces respectively. Here, the impurity spins are treated as classical, with a locally varying direction determined by the spin vector $\mathbf{S}(\mathbf{r}) = (\sin\theta_{\mathbf{r}} \cos\phi_{\mathbf{r}}, \sin\theta_{\mathbf{r}} \sin\phi_{\mathbf{r}}, \cos\theta_{\mathbf{r}})$, with $|\mathbf{S}(\mathbf{r})| = 1$. We set the angles between adjacent spins as follows: $\theta_{\mathbf{r}} = \pi/2$ and $\phi_{\mathbf{r}} = \mathbf{g} \cdot \mathbf{r} = (g_{x} x + g_{y} y)$, which generates a spin-spiral configuration. Here, $\mathbf{g}$ is the pitch vector, which determines the period and propagation direction of the spin spiral.

Considering a unitary transformation $U = e^{-\frac{i}{2} \phi_{\mathbf{r}} \sigma_{z}}$, the model can be mapped onto an effective model with spin-orbit coupling (SOC) in the presence of a Zeeman field~\cite{2D_spin_tex4,2D_spin_tex4,2D_spin_tex5}. Applying this unitary transformation and transforming to reciprocal space, we obtain the following continuum model in momentum space as,
\begin{eqnarray}\label{Conti_momentum_static_Ham}
	\tilde{H}(\mathbf{k}) &=& \xi_{\mathbf{k}} \tau_{z} +   \frac{1}{2} (\mathbf{g} \cdot \mathbf{k})  \sigma_{z} \tau_{z} + J \sigma_{x}  + \Delta_{0} \tau_{x}\ ,
\end{eqnarray}
where, $\xi_{\mathbf{k}} = \frac{1}{2} (\mathbf{k}^{2} + \mathbf{g}^{2}) - \mu$ and the second term representing an effective SOC. Here, $J$ acts as an effective Zeeman field along the $x$-direction.

In order to calculate the topological invariant of the system, we construct the lattice regularized model by substituting $k_{x,y} \rightarrow \sin{k_{x,y}}$ 
and $[1 - (k_{x,y}^{2}/2)] \rightarrow \cos{k_{x,y}}$~\cite{2D_spin_tex4} in the low energy continuum model (i.e., in Eq.~(\ref{Conti_momentum_static_Ham})) and obtain,  
\begin{eqnarray}\label{TB_momentum_static_Ham}
	H_{\text{L}}(\mathbf{k}) &=& \epsilon_{\mathbf{k}} \tau_{z} +  \frac{1}{2} (g_{x} \sin k_{x} + g_{y} \sin k_{y}) \tau_{z} \sigma_{z} \non \\
	&& + J \sigma_{x} + \Delta_{0} \tau_{x} \ ,
\end{eqnarray}
with $\epsilon_{\mathbf{k}} = (2 - \cos k_{x} - \cos k_{y}) + \frac{1}{2}(g_{x}^{2} + g_{y}^{2}) - \mu $ and other parameters are defined as earlier. While the presence of Zeeman term breaks the time-reversal symmetry (TRS), the Hamiltonian preserves chiral symmetry, \ie $\Sigma^{-1} H_{\text{L}}(\mathbf{k}) \Sigma = - H_{\text{L}}(\mathbf{k})$ with the chiral symmetry operator $\Sigma = \tau_{y} \sigma_{z}$.

To study the given 2D system in a real-space geometry with open boundary conditions (OBCs) along both directions, we rewrite the model Hamiltonian discussed in Eq.~(\ref{Conti_real_static_Ham}) in a tight-binding form. Considering the Nambu basis $\Psi_{i} = (c_{i \uparrow}, c_{i \downarrow}, c_{i \downarrow}^{\dagger}, -c_{i \uparrow}^{\dagger})^{\text{T}}$, the real-space tight-binding model of the 2D Shiba lattice can be written as~\cite{2D_spin_tex1,2D_spin_tex4},
\begin{eqnarray}\label{TB_real_static_Ham}
	H &=& \sum_{i,j}  \Psi_{i}^{\dagger} \big[ (-t_{ij} -  \mu \delta_{ij}) \Gamma_{1} + \delta_{ij} (J S_{i}^{x} \Gamma_{2}  \non \\
	&&+ J S_{i}^{y} \Gamma_{3} + \Delta_{0} \Gamma_{4})\big]  \Psi_{j} \ ,
\end{eqnarray}
Here, $c_{i,\alpha}^{\dagger}$ ($c_{i,\alpha}$) denotes the creation (annihilation) operator of an electron at site $i$ with spin $\alpha \, (= \uparrow, \downarrow)$. The parameter $t_{ij} = t$ corresponds to nearest-neighbor hopping amplitude. The other parameters $\mu$, $\Delta_{0}$, and $J$ represent the onsite chemical potential, the proximity-induced $s$-wave superconducting pairing strength, and the exchange coupling, respectively. The third term describes a magnetic texture, where the spin at each site is oriented along the unit vector with components ${S}_{i}^{x} = \sin \theta_{i} \cos \phi_{i}$ and ${S}_{i}^{y} = \sin \theta_{i} \sin \phi_{i}$. The angles $\theta_{i}$ and $\phi_{i}$ vary in the same manner as defined earlier in the context of the continuum model. 
We also define $\Gamma_{1} = \tau_{z} \sigma_{0}$, $\Gamma_{2} = \tau_{0} \sigma_{x}$, $\Gamma_{3} = \tau_{0} \sigma_{y}$, and $\Gamma_{4} = \tau_{x} \sigma_{0}$.

\begin{figure*}[t]
	\centering
	\subfigure{\includegraphics[width=1\textwidth]{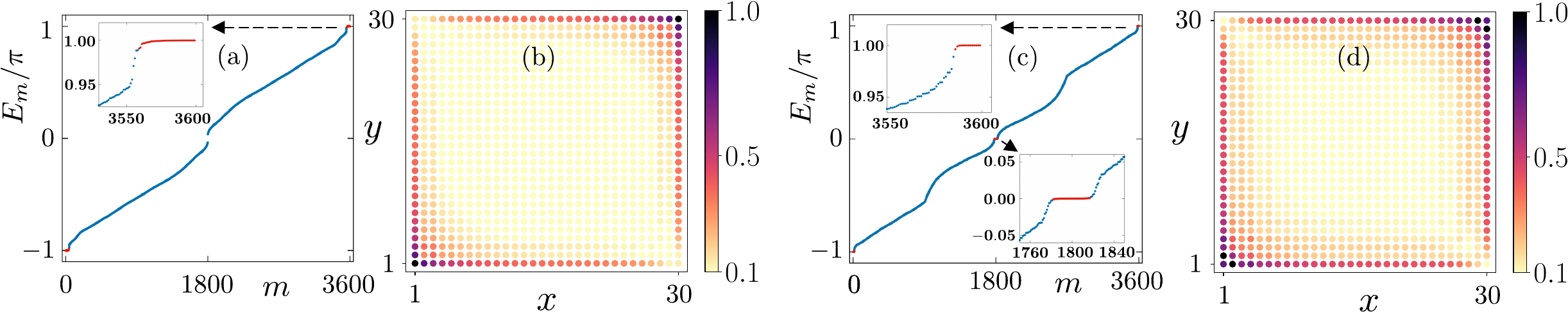}}
	\caption{In panels~(a) and (c), we depict the quasi-energy spectra as a function of the state index $m$ for the driven system, choosing the driving frequency $\Omega = 5\Delta_{0}$ and $\Omega = 10\Delta_{0}$ respectively. The red color dots indicate $0$- and $\pi$-FMFEMs. In panel~(b), we illustrate the LDOS distribution of $\pi$-FMFEMs (corresponding to the panel (a)) localized at the edges of the 2D domain. 
	Similarly, in panel~(d) we illustrate the LDOS spectra corresponding to the $0$-FMFEMs in panel~(c). The other model parameters are chosen as $\Delta_{0}=t=1$, $J = 2\Delta_{0}$, $V = 5\Delta_{0}$, $g_{x}=g_{y}=\pi / 2$.
	}
	\label{Fig2}
\end{figure*}
\subsection{Driving Protocol and Formalism}
In order to realize the FMFEMs, here we consider a time-dependent sinusoidal drive in the onsite chemical potential as following,
\begin{eqnarray}\label{chemical_drive}
	V(t) &=& V \cos (\Omega t ) \Gamma_{1}\ ,
\end{eqnarray}
where, $V$ and $\Omega \, (= 2\pi/T)$ represent the amplitude and frequency of the harmonic drive with time period $T$. Since $V(t)$ is time-periodic, the full Hamiltonian $\mathcal{H}(t) = H + V(t)$ is also periodic,\ie $\mathcal{H}(t+T) = \mathcal{H}(t)$. Here, $H$ corresponds to the static part of the Hamiltonian, as given in Eq.~(\ref{TB_real_static_Ham}) (real space) and Eq.~(\ref{TB_momentum_static_Ham}) (momentum space). Although we discuss a few perturbative techniques later to analyse our  system, here we primarily focus on the time-domain formalism. The time-evolution operator $U(t,t_{0})$ can be written as the following time-ordered product~\cite{2D_Floq_TSC6},
\begin{eqnarray}\label{time_evol_op}
	U(t,0) &=& \mathcal{T} \exp [-i \int_{0}^{t} dt^{\prime} \mathcal{H}(t^{\prime})] \non \\
	&=&  \exp [-i \sum_{j=0}^{N-1} \mathcal{H}(t_{j}) \Delta t] \non \\
	&=& \prod_{j=0}^{N-1}  \exp [-i  \mathcal{H}(t_{j}) \Delta t] \non \\
	&=& \prod_{j=0}^{N-1}  U(t_{j} + \Delta t,t_{j})
\end{eqnarray}
where, $\mathcal{T}$ denotes the time ordering. Trotter decomposition~\cite{2D_Floq_TSC6} is employed to express the exponential of a sum as a product of exponentials. For a full driving cycle, the time-evolution operator is referred to as the Floquet operator, which can be constructed from Eq.~(\ref{time_evol_op}) by substituting $t$ with $T$, i.e., $U(T,0)$. The Floquet operator ($U(T,0)$) plays the pivotal role in a driven system and allows 
us to calculate the quasi-energy spectrum $E_{m} \in [-\pi, \pi]$. In the following sections, we analyze the eigenvalue spectra, local density of states (LDOS), and the calculation of topological invariants within this formalism.


\section{Generation of anomalous FMFEM\lowercase{s} and their time evolution}\label{Sec:III}
In this section, we present all the numerical results obtained from the real-space tight-binding model using the Floquet operator and the time-evolution operator discussed earlier.

\subsection{Generation of anomalous FMFEMs}
Here, we discuss the results based on the real-space model Hamiltonian in Eq.~(\ref{TB_real_static_Ham}), with a drive in the chemical potential as given in Eq.~(\ref{chemical_drive}). The corresponding Floquet operator is constructed following Eq.~(\ref{time_evol_op}) and numerically diagonalized to obtain the quasi-energy spectrum and LDOS, as shown in Fig.~\ref{Fig2}. Starting from a topological phase of the static Hamiltonian ($J = 2\Delta_{0}$), we periodically drive the system. For two different driving frequencies, $\Omega = 5\Delta_{0}$ and $\Omega = 10\Delta_{0}$, we present the eigenvalue spectra in Fig.~\ref{Fig2}(a) and Fig.~\ref{Fig2}(c), respectively. We find that at low driving frequency (compared to the static system's band-width energy scale $\sim 10t$ obtained from effective momentum space spectra~\cite{2D_spin_tex4}), the system hosts $\pi$-FMFEMs, as shown in red in Fig.~\ref{Fig2}(a), while zero-quasienergy states are absent. The corresponding LDOS is presented in Fig.~\ref{Fig2}(b). This exhibits the 1D localized $\pi$-edge modes. However, in the intermediate-frequency regime ($\Omega = 10\Delta_{0}$), we observe the coexistence of both $0$- and $\pi$-FMFEMs, as shown in Fig.~\ref{Fig2}(c), with the insets highlighting the regions where the $0$- and $\pi$-modes appear. The corresponding LDOS for the $0$-FMFEMs is shown in Fig.~\ref{Fig2}(d). Since there is no static counterpart of the $\pi$-FMFEMs, they are often referred to as anomalous FMFEMs. We note that although the results are shown for a specific set of parameter values, similar features persist over a range of finite parameter values. 
\subsection{Time dynamics of the FMFEMs}
Here, we discuss the time evolution of the Floquet states to understand how the $0$- and $\pi$-Floquet modes appear and evolve with time. To this end, we consider $U(t,0)$ at time $t$ and diagonalize it to obtain the time-dependent quasi-energy spectrum $E(t)$. 

\begin{figure*}[]
	\centering
	\subfigure{\includegraphics[width=0.8\textwidth]{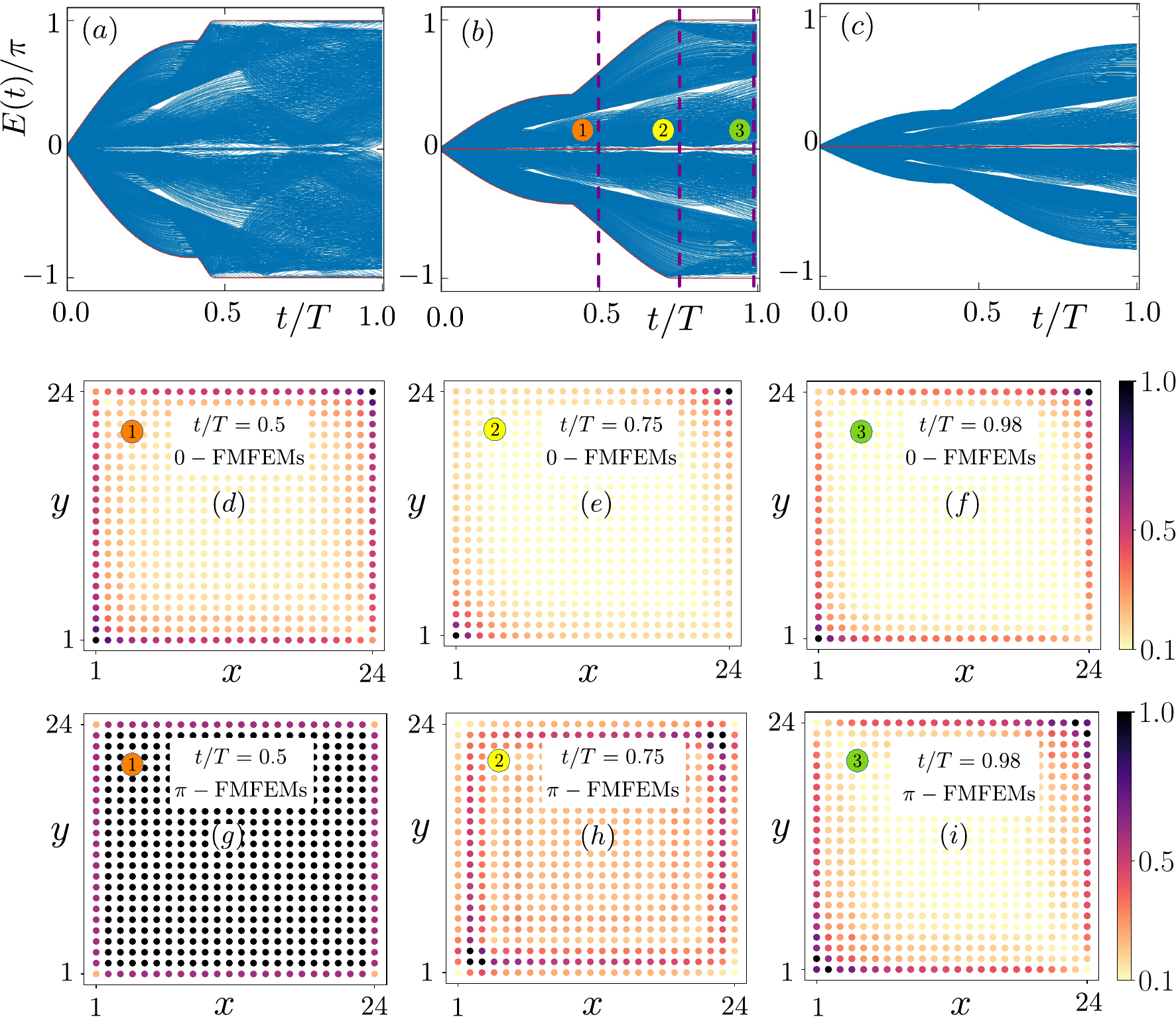}}
	\caption{
	In panels (a), (b), and (c), we demonstrate the time-dependent eigenvalue spectra $E(t)$ of the time-evolution operator $U(t,0)$ as a function of time $t$ over a full driving period choosing $\Omega = 5 \Delta_{0}$, $10 \Delta_{0}$, and $15\Delta_{0}$, respectively. Panels (d)–(f) show the evolution of the 
	edge states corresponding to the $0$-FMFEMs at intermediate frequency ($\Omega = 10 \Delta_{0}$), considering time scales $t = 0.5T$, $0.75T$, and $0.98T$, respectively. The same is depicted for the $\pi$-FMFEMs in panels~(g)–(i) considering the same time instants mentioned above. The other model parameters are chosen as $\Delta_{0} = t = 1$, $J = 2\Delta_{0}$, $V = 5\Delta_{0}$, and $g_{x} = g_{y} = \pi/2$.
	}
	\label{Fig3}
\end{figure*}

In Figs.~\ref{Fig3}(a)-(c), we depict the time evolution of the quasi-energy spectrum for three different values of the driving frequency, $\Omega = 5\Delta_{0}, 10\Delta_{0}, 15\Delta_{0}$, respectively. We find that in the low-frequency regime, the $\pi$-modes appear before half the driving period and persist up to $t = T$, while the zero-energy states disappear at $t = T$. In the intermediate-frequency regime, the $\pi$-modes emerge near $t = 0.75T$, while both the $\pi$- and $0$-modes persist at $t = T$, as shown in Fig.~\ref{Fig3}(b). Finally, in the high-frequency regime, the $\pi$-modes do not appear, and only the $0$-modes remain, as shown in Fig.~\ref{Fig3}(c). The latter is the characteristic of zero-photon sector in the high frequency regime. For the intermediate-frequency case, we present the evolution of the LDOS at three different time scales considering both the $0$- and $\pi$-FMFEMs. In Fig.~\ref{Fig3}(b), the time instants at which the LDOS is calculated, $t = 0.5T, 0.75T, 0.98T$, are marked by red, yellow, and green circles, respectively. The corresponding evolution of the LDOS distribution be in tune with the zero-energy FMFEMs is shown in Figs.~\ref{Fig3}(d)-(f). Similarly, the evolution of the LDOS spectra corresponding to the $\pi$-FMFEMs is displayed in Figs.~\ref{Fig3}(g)-(i). This clearly manifests the edge localization of the corresponding $0$- and $\pi$-FMFEMs at different time instants.

\section{Topological Characterization}\label{Sec:IV}

\begin{figure*}[]
	\centering
	\subfigure{\includegraphics[width=0.8\textwidth]{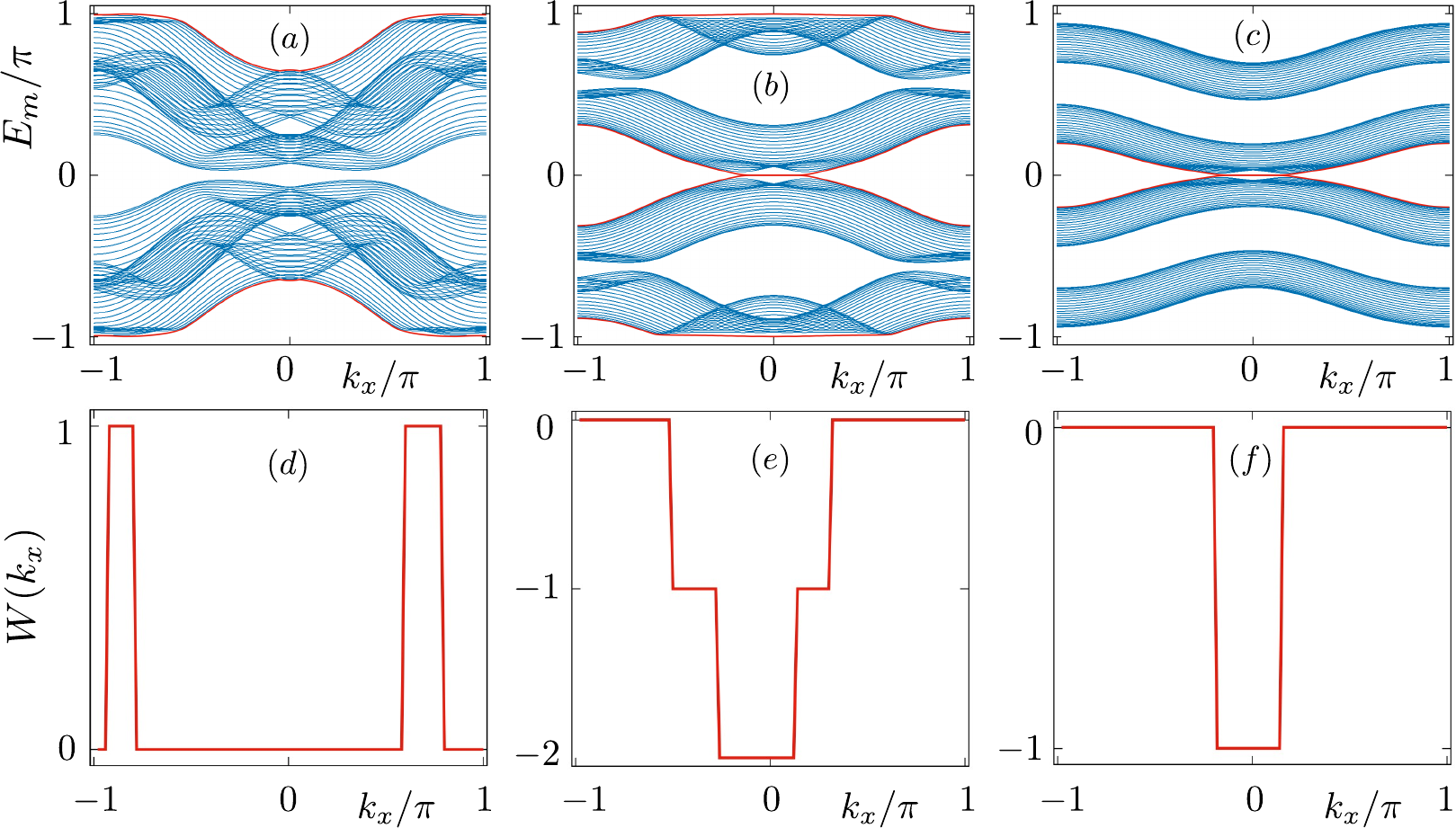}}
	\caption{Panels (a)-(c) correspond to the quasi-energy edge spectra shown as a function of momentum $k_{x}$ for three different values of the driving frequency, $\Omega = 5\Delta_{0}$, $10\Delta_{0}$, and $15\Delta_{0}$, respectively. On the other hand, panels (d)-(f) exhibit the corresponding winding number as a function of momentum $k_{x}$ for the same choice of respective driving frequencies. The other model parameters are chosen as follows: $V = 5\Delta_{0}$, $J = 2\Delta_{0}$, $\Delta_{0} = t = 1$, and $g_{x} = g_{y} = \pi/2$.
	}
	\label{Edge_and_invariants}
\end{figure*}
In this section, we perform the topological characterization of the 2D Shiba lattice based on the bulk Hamiltonian (see Eq.~(\ref{TB_momentum_static_Ham})) driven by the same harmonic drive (see Eq.~(\ref{chemical_drive})). Note that, as the static Hamiltonian exhibits a gapless bulk spectrum in the topological regime~\cite{2D_spin_tex4}, the Chern number is not well defined in this region. Instead, the topological phase can be characterized by momentum-dependent winding numbers~\cite{2D_SOC_Bz3}, since the Hamiltonian satisfies the chiral symmetry.

To begin with, we consider the static Hamiltonian in the chiral basis. The chiral symmetry operator for the static Hamiltonian is given by $\Sigma = \tau_{y} \sigma_{z}$ and satisfy $\Sigma^{-1} H_{\text{L}}(\mathbf{k}) \Sigma = - H_{\text{L}}(\mathbf{k})$. In the chiral basis the Hamiltonian turns into block off-diagonal form as given below, 
\begin{eqnarray}\label{Static_chiral_symmetry}
	U_{\Sigma}^{\dagger} H_{L}(\mathbf{k}) U_{\Sigma} &=
	\left( \begin{array}{cc}
		0  & q_{+}(\mathbf{k}) \\
		q_{-}(\mathbf{k}) & 0
	\end{array}\right)\ ,\
\end{eqnarray}
From the knowledge of $q_{\pm}(\mathbf{k})$ one can calculate the winding number for the static system. 

In presence of drive the definition of the chiral symmetry is generalized in the following way,
\begin{eqnarray}\label{Floq_chiral_symmetry1}
	\Sigma^{-1} H_{\text{L}}(t,\mathbf{k}) \Sigma = - H_{\text{L}}(-t,\mathbf{k})\ .
\end{eqnarray} 
This imposes one of the following constraints on the time-dependent part of the Hamiltonian: either $\Sigma^{-1} V(t,\mathbf{k}) \Sigma = -V(t,\mathbf{k})$ with $V(t,\mathbf{k}) = +V(-t,\mathbf{k})$, or $\Sigma^{-1} V(t,\mathbf{k}) \Sigma = +V(t,\mathbf{k})$ with $V(t,\mathbf{k}) = -V(-t,\mathbf{k})$, must be satisfied in order to preserve the chiral symmetry~\cite{CS1,CS2}. In our case, the first condition is satisfied, which corresponds to an extension of the chiral symmetry for the dynamic case. We then compute the winding number using the evolution operator, which serves as the dynamical analog of the Hamiltonian. With this, we express the evolution operator in the chiral basis as,
\begin{eqnarray}\label{Floq_chiral_symmetry2}
	U_{\Sigma}^{\dagger} U(\mathbf{k},T) U_{\Sigma} &=
	\left( \begin{array}{cc}
		0  & u_{+}(\mathbf{k}) \\
		u_{-}(\mathbf{k}) & 0
	\end{array}\right)\ ,\
\end{eqnarray}
Then using the block off-diagonal forms we compute the dynamical winding number as,
\begin{eqnarray}\label{Floq_winding}
	W(k_{x}) = \Big| \pm \frac{i}{2\pi} \int_{-\pi}^{\pi} dk_{y} \text{Tr}\big[ \lbrace{u_{\pm}(\mathbf{k})\rbrace}^{-1} \partial_{k_{y}} u_{\pm}(\mathbf{k})\big] \Big|\ .
\end{eqnarray}

On the other hand, to compare with the winding number, we also calculate the edge spectra by imposing periodic boundary condition (PBC) along the $x$-direction and OBC along the $y$-direction. Considering this edge Hamiltonian with a drive in the chemical potential according to Eq.~(\ref{chemical_drive}), we examine the edge spectra for three different driving frequencies.

In the low-frequency regime,\ie for $\Omega = 5\Delta_{0}$ (see Fig.~\ref{Edge_and_invariants}(a)), we observe only the appearance of $\pi$-FMFEMs near the $k_{x} = \pm \pi$ regions, while no $0$-FMFEMs are present in this regime. This is consistent with our real-space (employing OBC in both directions) quasi-energy spectra, which also exhibit only $\pi$-modes, as presented in Fig.~\ref{Fig2}(a). In the intermediate-frequency regime (i.e. $\Omega = 10\Delta_{0}$), both the $0$- and $\pi$-FMFEMs appear in the edge spectra, as shown in Fig.~\ref{Edge_and_invariants}(b). This again agrees well with the real-space results shown in Fig.~\ref{Fig2}(b). Finally, in the high-frequency regime (i.e. $\Omega = 15\Delta_{0}$), depicted in Fig.~\ref{Edge_and_invariants}(c), only the $0$-FMFEMs are present which is again consistent with the edge state spectra displayed in Fig.~\ref{Fig2}(c). 

Each of the edge spectra is associated with the winding number shown as a function of $k_{x}$ for the corresponding frequency regimes, in Figs.~\ref{Edge_and_invariants}(d)-(f). 
Note that, the winding number acquires finite integer values in the presence of $0$- or $\pi$-FMFEMs. In the low-frequency regime, the winding number takes the value $+1$, indicating the presence of only $\pi$-modes. In the intermediate frequency regime, it acquires the value $-2$ in regions where both $0$- and $\pi$-FMFEMs coexist. 
Finally, in the high-frequency regime, the winding number is consistent (value takes -1) with the presence of only $0$-FMFEMs.

\begin{figure}[]
	\centering
	\subfigure{\includegraphics[width=0.5\textwidth]{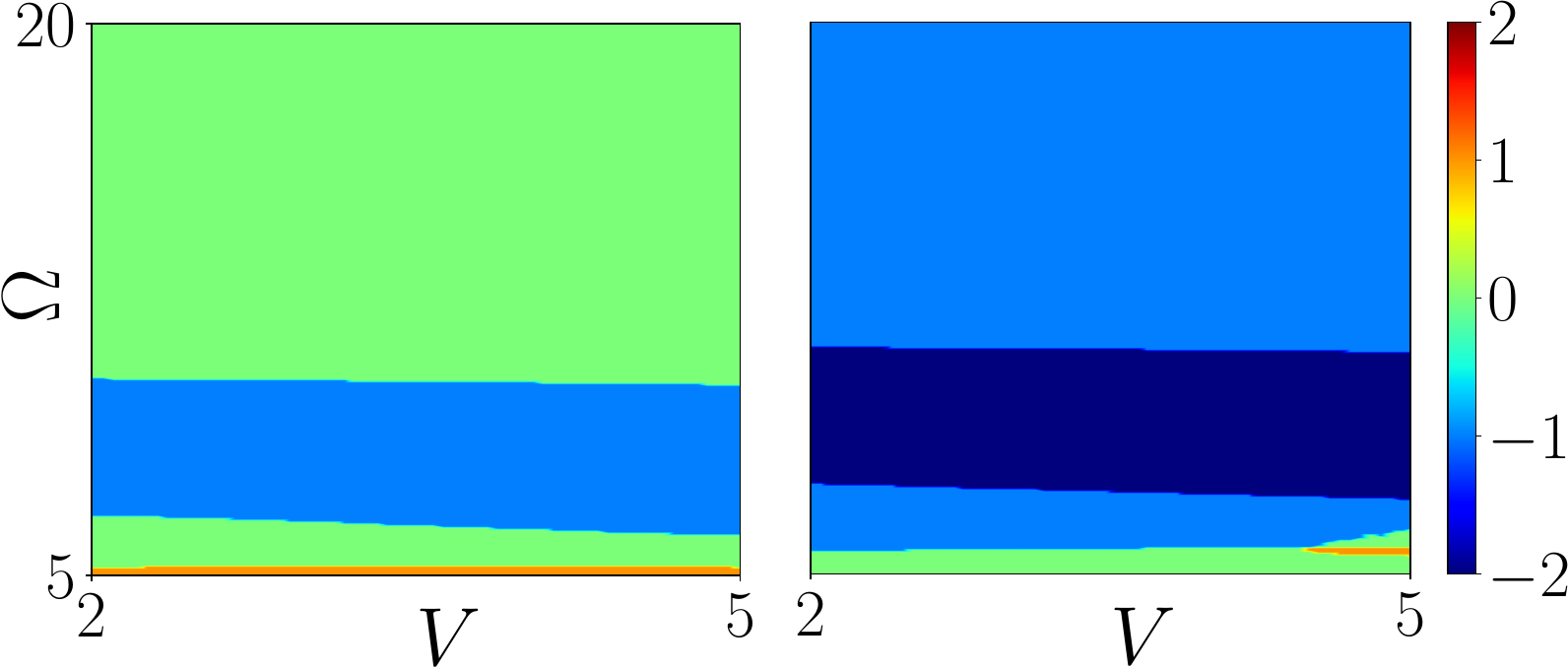}}
	\caption{Density plots of the winding number (evaluated at $k_{x} = 0$) is displayed in the plane of driving amplitude ($V$) and frequency ($\Omega$). We present the phase diagram for two different values of the exchange coupling: $J = 1.6 \Delta_{0}$ (panel~(a)) and $J = 2 \Delta_{0}$ (panel~(b)), corresponding to the trivial and topological phases of the static system, respectively. The other model parameters are chosen as $\Delta_{0} = t = 1$, $V = 5\Delta_{0}$, and $g_{x} = g_{y} = \pi/2$.
	}
	\label{Phase_diagram}
\end{figure}

To better understand the effect of the driving parameters on the topological phases, we construct phase diagrams of the driven system starting from the corresponding topologically trivial and non-trivial phases of the static system. In Fig.~\ref{Phase_diagram}, we depict the winding number in the plane of the driving amplitude ($V$) and frequency ($\Omega$). However, in our calculation, the winding number is evaluated as a function of momentum ($k_{x}$). Therefore, 
to construct the phase diagram in the $V$--$\Omega$ plane, we consider the winding number only at $k_{x} = 0$ (i.e. $W(0)$). Although, this approach does not provide the complete phase diagram, it nevertheless captures several distinct phases. 

In Fig.~\ref{Phase_diagram}(a) and Fig.~\ref{Phase_diagram}(b), we present the $k_{x}=0$ phase diagrams for $J = 1.6 \Delta_{0}$ (corresponding to the trivial phase of the static system) and $J = 2 \Delta_{0}$ (corresponding to the non-trivial TSC phase of the static system), respectively. It can be seen that while 
$J = 1.6 \Delta_{0}$ corresponds to a trivial phase in the static system, the driven system exhibits non-trivial Floquet TSC phases with finite winding number. On the other hand, the static topological phase undergoes multiple topological phase transitions in the presence of the harmonic drive manifesting higher Chern numbers with multiple FMFEMs.

\section{Perturbative Analysis}\label{Sec:V}

The exact analytical calculation of Floquet Hamiltonian for a generic driven quantum system is not possible as it is an infinite dimensional matrix in frequency space. Hence, various perturbative schemes are performed to obtain the effective Floquet Hamiltonian~\cite{FM,BW1,FPT1}. In this section we discuss the construction of an effective Hamiltonian using the Brillouin-Wigner (BW) and Floquet perturbation theory (FPT) and compare the quasi-energy spectra with the exact numerical results.

Before we deal with the details, here we rewrite the real space tight binding Hamiltonian (Eq.~(\ref{TB_real_static_Ham})) in presence of the Harmonic drive in a compact form as,
\begin{eqnarray}\label{time_dependent_Hamiltonian}
	H(t) &=& A_{1} \Gamma_{1} + A_{2} \Gamma_{2} + A_{3} \Gamma_{3} + A_{4} \Gamma_{4} + V \cos(\Omega t ) \Gamma_{1}\ , \non \\
\end{eqnarray}
These gamma matrices are already defined earlier and the corresponding coefficients $A_{1}$, $A_{2}$,
$A_{3}$ and $A_{4}$ can be easily identified from Eq.~(\ref{TB_real_static_Ham}) and Eq.~(\ref{chemical_drive}).

\begin{figure*}[t]
	\centering
	\subfigure{\includegraphics[width=0.9\textwidth]{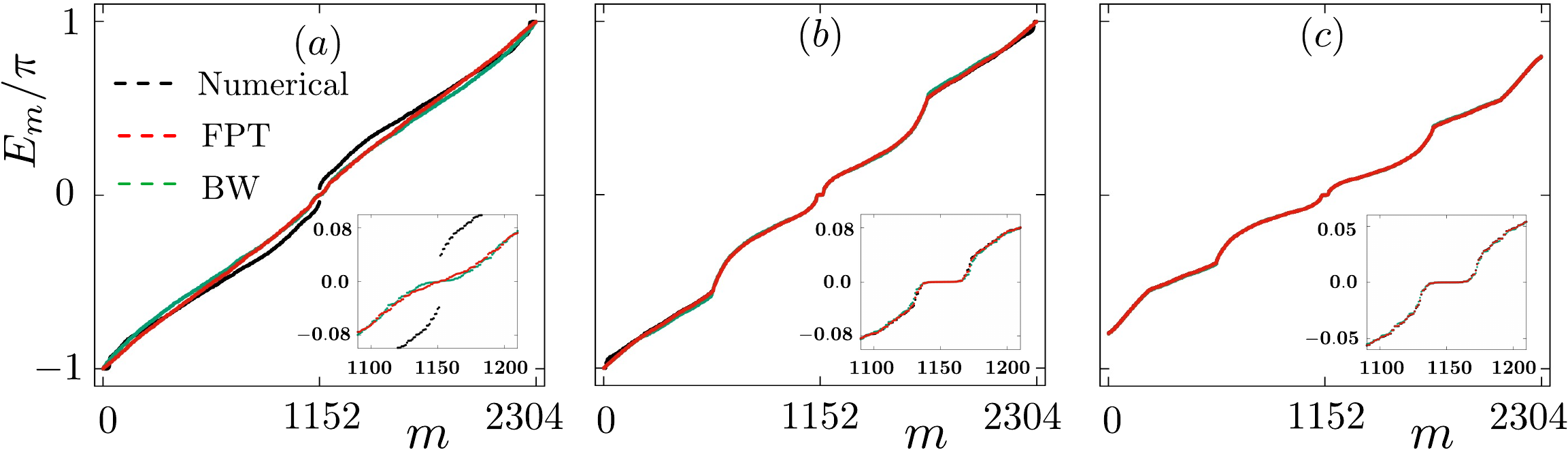}}
	\caption{In panels (a), (b) and (c), we compare the quasi-energy spectra obtained from the perturbative schemes with those obtained numerically from the exact Floquet operator considering different frequency regime. The green, red, and black curves represent the results acquired from BW, FPT, and exact numerical calculations, respectively. In panels~(a), (b) and (c), the results are shown for $\Omega = 5\Delta_{0}$ (low frequency), $\Omega = 10\Delta_{0}$ (intermediate frequency), and $\Omega = 15\Delta_{0}$ (high frequency), respectively. To obtain these results, the driving amplitude is fixed at $V = 5\Delta_{0}$. The other model parameters are chosen as $\Delta_{0} = t = 1$ and $g_{x} = g_{y} = \pi/2$.
	}
	\label{Peturbation}
\end{figure*}

\subsection{Brillouin-Wigner Perturbation Theory}
BW perturbation theory has been successfully employed to obtain an effective Hamiltonian in a periodically driven quantum system in the high frequency limit~\cite{BW1,BW2,BW3}. Utilizing the time-periodicity of the Hamiltonian it is possible to write an effective Hamiltonian in terms of the Fourier components and in powers of $1/\Omega$ as~\cite{BW1},
\begin{eqnarray}\label{BW1}
	H_{\text{BW}}^{\text{eff}} &=& \sum_{n=0}^{\infty} H_{\text{BW}}^{(n)}\ ,
\end{eqnarray} 
where the first few order terms can be written as,
\begin{eqnarray}\label{BW2}
	H_{\text{BW}}^{(0)} &=& H_{0,0}\ , \non \\
	H_{\text{BW}}^{(1)} &=& \sum_{n \neq 0}\frac{H_{0,n} H_{n,0}}{n \Omega}\ , \non \\
	H_{\text{BW}}^{(2)} &=& \sum_{n,m \neq 0}\frac{H_{0,n} H_{n,m} H_{m,0}}{n m\Omega^{2}} - \frac{H_{0,n} H_{n,0} H_{0,0}}{n^{2} \Omega^{2}}\ ,
\end{eqnarray} 
In our calculation, we consider contribution arising upto $\mathcal{O}(1/\Omega^{2})$ in the perturbative expansion. As the higher order contributions are vanishingly small we ignore those terms. The Fourier components ($H_{n,m}$) in the above are given by,
\begin{eqnarray}\label{BW3}
	H_{n,m} &=& \frac{1}{T} \int_{0}^{T} dt H(t) e^{i (n-m) \Omega t}\ .
\end{eqnarray}

Following the driving protocol we have, $V(t) = V \cos(\Omega t) \Gamma_{1}$ and it can be shown that, 
\begin{eqnarray}\label{BW4}
 	H_{0,n} &=& \frac{V}{2} \Gamma_{1} \delta_{n,\pm 1}\ ,
\end{eqnarray} 
Using Eq.~(\ref{BW3}), Eq.~(\ref{BW4}) and our model Hamiltonian we find that, 
\begin{eqnarray}\label{BW5}
	H_{\text{BW}}^{(0)} = H, \hspace{7pt}
	H_{\text{BW}}^{(1)} = \frac{V^{2}}{4 \Omega} I, \hspace{7pt}
	H_{\text{BW}}^{(2)} = -\frac{V^{2}}{2 \Omega} A_{4} \Gamma_{4}\ ,
\end{eqnarray} 
Therefore the effective BW Hamiltonian can be written as,
\begin{eqnarray}\label{BW6}
 	H^{\text{eff}}_{\text{BW}} &=& A_{1} \Gamma_{1} + A_{2} \Gamma_{2} + A_{3} \Gamma_{3} + A_{4} \left(1 -\frac{V^{2}}{2 \Omega}\right) \Gamma_{4}\ , 
\end{eqnarray}


\subsection{Floquet Perturbation Theory}
FPT allows one to compute the Floquet operator perturbatively~\cite{FPT1, FPT2}, in comparison to the previous method where the Floquet Hamiltonian is calculated perturbatively. The time dependent Hamiltonian given by Eq.~(\ref{time_dependent_Hamiltonian}) can be written in terms of the time-dependent part $V(t)$ and time-independent part $H$ as, 
\begin{eqnarray}\label{FPT1}
	H(t) &=& V(t) + H\ , 
\end{eqnarray} 
Under this perturbation theory, the time dependent part $V(t)$ is required to be exactly solvable, and time-independent part $H$ is treated perturbatively. 
Also, the perturbation parameter in this scheme is the ratio of amplitudes in the two parts of Eq.~(\ref{FPT1}), hence in the high driving amplitude limit
it works very well~\cite{FPT1, FPT2}.  
\begin{eqnarray}\label{FPT2}
	U_{0}(t,0) &=& \mathcal{T} \exp [-i \int_{0}^{t} dt^{\prime} V(t^{\prime})] \non \\
	&=& \exp \left[- \frac{iV}{\Omega} \sin(\Omega t) \Gamma_{1}\right]\ .
\end{eqnarray} 

Under the interaction picture, the time evolution operator for $H(t)$ can be written as~\cite{FPT1, FPT2, 2D_Floq_TSC6, 1D_Floq_TSC6}, 
\begin{eqnarray}\label{FPT3}
	U_{I}(t,0) &=& I + (-i) \int_{0}^{t} dt^{\prime} H_{I}(t^{\prime}) \non \\ 
	&+& (-i)^{2} \int_{0}^{t} dt^{\prime} H_{I}(t^{\prime}) \int_{0}^{t^{\prime}} dt^{\prime\prime} H_{I}(t^{\prime\prime}) + ... \non \\   
	&=& I + U_{I}^{(1)} (t,0) + U_{I}^{(2)}(t,0) + ...\ ,
\end{eqnarray} 
with,
\begin{eqnarray}\label{FPT4}
	H_{I}(t) &=& U_{0}^{\dagger}(t,0) H U_{0}(t,0)\ ,
\end{eqnarray} 
Finally the full time-evolution operator can be written as,
\begin{eqnarray}\label{FPT5}
	U_{\text{P}}(t,0) &=& U_{0}(t,0) U_{I}(t,0)\ .
\end{eqnarray} 

In order to calculate the $H_{I}(t)$, we note that from Eq.~(\ref{FPT4}),
\begin{eqnarray}\label{FPT6}
 	H_{I}(t) &=& e^{i M \Gamma_{1}} H e^{- i M \Gamma_{1}} \ ,
\end{eqnarray} 
where, $M = \frac{V}{\Omega} \sin(\Omega t)$. Using the commutation and anti-commutation relations between the $\Gamma$- matrices \ie $[\Gamma_{1},\Gamma_{2}] = [\Gamma_{1},\Gamma_{3}] = 0, \lbrace\Gamma_{1},\Gamma_{4} \rbrace = 0$, we obtain,

\begin{eqnarray}\label{FPT7}
	e^{i M \Gamma_{1}} \Gamma_{j} e^{- i M \Gamma_{1}}  &=&  \Gamma_{j}, \hspace{5pt}\text{for} \hspace{5pt} j =1,2,3 \ ,\non \\
	e^{i M \Gamma_{1}} \Gamma_{4} e^{- i M \Gamma_{1}}  &=& e^{2 i M \Gamma_{1}} \Gamma_{4} \ ,
\end{eqnarray} 
These relations yield,
\begin{eqnarray}\label{FPT8}
	H_{I}(t) &=&  A_{1} \Gamma_{1} + A_{2} \Gamma_{2} + A_{3} \Gamma_{3} +  e^{2iM\Gamma_{1}} A_{4} \Gamma_{4}\ ,
\end{eqnarray} 

Following the Jacobi-Anger identity, $e^{i z \sin x} =  J_{0}(z) + \sum_{\lbrace n \rbrace \neq 0} J_{n}(z) e^{i n z}$ we simplify the above expression as,
\begin{eqnarray}\label{FPT9}
	H_{I}(t) &=&  A_{1} \Gamma_{1} + A_{2} \Gamma_{2} + A_{3} \Gamma_{3} +  \Big[ J_{0} \Big( \frac{2V}{\Omega} \Big)\non \\ 
	&+& \sum_{\lbrace n \rbrace \neq 0} J_{n} \left( \frac{2V}{\Omega} \Big) e^{i n \Omega t}  \right] A_{4} \Gamma_{4}\ ,
\end{eqnarray} 
To get the Floquet operator we substitute $t \rightarrow T$ in Eq.~(\ref{FPT5}) and obtain,
\begin{eqnarray}\label{FPT10}
	U_{\text{P}}(T,0) &=& U_{0}(T,0) U_{I}(T,0)\ ,
\end{eqnarray} 
Then employing $U_{0}(T,0) = I$ and Eq.~(\ref{FPT3}) we obtain,
\begin{eqnarray}\label{FPT11}
	U_{P}(T,0) &=& I + (-i) \int_{0}^{T} dt^{\prime} H_{I}(t^{\prime}) \non \\ 
	&+& (-i)^{2} \int_{0}^{T} dt^{\prime} H_{I}(t^{\prime}) \int_{0}^{t^{\prime}} dt^{\prime\prime} H_{I}(t^{\prime\prime}) + ... \non \\   
	&=& I + U_{P}^{(1)} (T,0) + U_{P}^{(2)}(T,0) + ...\ ,
\end{eqnarray} 

Considering the leading order term in the perturbation series we obtain the $U_{P}^{(1)} (T,0)$ as,
\begin{eqnarray}\label{FPT12}
	U_{P}^{(1)} (T,0) &=& (-i) \Big[A_{1} \Gamma_{1} + A_{2} \Gamma_{2} + A_{3} \Gamma_{3} \non \\
	&+& J_{0} \left(\frac{2V}{\Omega} \right)
	 A_{4} \Gamma_{4} \Big] T\ ,
\end{eqnarray} 

Therefore the effective Hamiltonian can be written as,
\begin{eqnarray}\label{FPT13}
	H^{\text{eff}}_{\text{FPT}} &=& A_{1} \Gamma_{1} + A_{2} \Gamma_{2} + A_{3} \Gamma_{3} 
	+ J_{0} \left( \frac{2V}{\Omega} \right)
	A_{4} \Gamma_{4}\ .
\end{eqnarray} 

In Fig.~\ref{Peturbation}, we depict the quasi-energy spectra obtained from the evolution operator [$U(T,0) = e^{-i H_{\text{eff}} T}$] employing the BW, FPT, and exact numerical calculations. To facilitate comparison, we use green, red, and black curves to represent the BW, FPT, and exact numerical results, respectively. In Figs.~\ref{Peturbation}(a)-(c), the results are shown for $\Omega = 5\Delta_{0}$ (low frequency), $\Omega = 10\Delta_{0}$ (intermediate frequency), and $\Omega = 15\Delta_{0}$ (high frequency), respectively, with the driving amplitude fixed at a relatively large value, $V = 5\Delta_{0}$. 
We find that while the perturbative results deviate from the exact results in the low-frequency regime, they exhibit good agreement in the intermediate- and high-frequency regimes (see Figs.~\ref{Peturbation}(b)-(c)). Although the $0$-FMFEMs can be understood analytically in certain limiting cases, the current perturbative approaches fail to capture the $\pi$-FMFEMs in the quasi-energy spectra. This is evident from Fig.~\ref{Peturbation}(a) where perturbative approaches exhibit gapless quasi-energy spectra around $0$-energy while it is gapped in our exact numerical calculation. Also $\pi$-FMFEMs appear neither in
BW nor in FPT while emergence of $\pi$-FMFEMs is evident from our exact numerical results (see also Fig.~\ref{Fig2}(a)). 



\section{Summary and Conclusions}\label{Sec:VI}

To summarize, in this article, we investigate the emergence of gapless Floquet topological superconducting phase based on a 2D spin texture deposited on the surface of a conventional $s$-wave superconductor (2D Shiba lattice) under a time-periodic harmonic drive, demonstrating the appearance of both $0$- and $\pi$-FMFEMs.
 
In the first part of our work, we study a real-space tight-binding model and analyze the quasi-energy spectrum obtained from the periodized evolution operator. Starting from a static topological phase, we drive the system at different frequencies. At low frequency, anomalous $\pi$-FMFEMs emerge while at intermediate frequency, both $0$- and $\pi$-FMFEMs coexist. We further compute the LDOS, confirming the presence of edge states for both the $0~\text{and}~\pi$-modes. Also, 
we examine the time evolution of the eigenvalue spectrum over one driving period at different frequency regime. For the intermediate frequency regime, we present LDOS at three different time scales, exhibiting the evolution of the edge states.

In order to calculate the topological invariants, we consider the momentum-space Hamiltonian obtained by a unitary transformation from the real space static Hamiltonian~\cite{2D_spin_tex4}.  Utilizing the chiral symmetry of the bulk Hamiltonian and extending it for the driven case, we calculate the dynamical winding number as a function of momentum from the evolution operator in the gapless Floquet TSC phase. On the other hand, to compare the winding number results, we calculate the edge spectra (by employing OBC in one direction and PBC in another direction) in the driven system. Considering the same parameter values as the real space lattice model, we obtain the quasi energy edge spectra for the low, intermediate and high frequency regime as a function of momentum. The results strongly support our findings based on the real-space tight-binding model. The Winding number takes finite integer values at regions where the $0$- and $\pi$- modes appear in the quasi energy edge spectra, indicating the topological nature. Moreover, we present a phase diagram of Winding number in the plane of driving amplitude and frequency, starting from the topological and non-topological regime of the static Hamiltonian. Finally, we perform a perturbative analysis to gain analytical insight into our numerical  results. We employ both the BW expansion and FPT, and compare their predictions with the exact numerical results obtained from the real-space tight-binding Hamiltonian across three driving-frequency regimes. We find that these perturbative approaches fail to capture the $\pi$-FMFEMs in different frequency regime (low and intermediate), but exhibit good agreement with the numerical results obtained in the intermediate- and high-frequency regimes as far as $0$-FMFEMs are concerned.
 
From the perspective of experimental feasibility, recent studies on the hybrid magnetic-superconducting platform Mn/Nb(110) have demonstrated the simultaneous presence of antiferromagnetism and superconductivity~\cite{Mn_Nb_expt1}. Motivated by these developments, one may envision an experimental architecture consisting of a monolayer of magnetic adatoms (such as Mn or Cr) deposited on the surface of a conventional $s$-wave superconductor like Nb or Al. Related heterostructures have already emerged as promising platforms for realizing topological superconducting phases and Majorana zero-energy 
excitations~\cite{Imp_chn_expt2,expt2,expt3}. Furthermore, a time-dependent sinusoidal driving of the local chemical potential can be implemented experimentally through the application of an ac gate voltage, to realize the Floquet TSC phase in such a 2D heterostructure.

\subsection*{Acknowledgments}
K.B. acknowledges Debashish Mondal for stimulating discussions. K.B. and A.S. acknowledge SAMKHYA: High-Performance Computing Facility provided by Institute of Physics, Bhubaneswar, and the two workstations provided by the Institute of Physics, Bhubaneswar from the DAE APEX project for numerical computations.

\subsection*{Data Availability Statement}
The datasets generated and analyzed during the current study are available from the corresponding author upon reasonable request.

\bibliography{bibfile}{}

\end{document}